\documentclass[iop, twocolumn, tighten]{emulateapj}
\usepackage{amsmath}
\usepackage{color}
\definecolor{redgray}{rgb}{0.0,0.0,0.0}
\definecolor{red}{rgb}{0.0,0.0,0.0}
\newcommand{\red}{\textcolor{red}}

\slugcomment{Submitted to ApJ on Dec.11, 2015; Revised on Feb.8, 2016; Accepted on Feb.23, 2016}
\shortauthors{Uehara et al.}
\shorttitle{Seven UOIs Detected by Visual Inspection}
\begin{document}
%%%%%%%%%%%%%%%%%%%%%%%%%%%%%%%%%%%%%%%%%%%%%%%%%%%%%%%%%%%%%%%%%%%%%%
\title{Transiting Planet Candidates Beyond the Snow Line Detected by Visual Inspection of 7557 {\it Kepler} Objects of Interest}
%%%%%%%%%%%%%%%%%%%%%%%%%%%%%%%%%%%%%%%%%%%%%%%%%%%%%%%%%%%%%%%%%%%%%%

%%%%%%%%%%%%%%%%%%%%%%%%%%%%%%%%%%%%%%%%%%%%%%%%%%%%%%%%%%%%%%%%%%%%%%
\author{Sho Uehara\altaffilmark{1}} 
\author{Hajime Kawahara\altaffilmark{2,3}} 
\author{Kento Masuda\altaffilmark{4}} 
\author{Shin'ya Yamada\altaffilmark{1}} 
\author{Masataka Aizawa\altaffilmark{4}} 
\altaffiltext{1}{Department of Physics, Tokyo Metropolitan University, 
Tokyo 192-0397, Japan}
\altaffiltext{2}{Department of Earth and Planetary Science, 
The University of Tokyo, Tokyo 113-0033, Japan}
\altaffiltext{3}{Research Center for the Early Universe, 
School of Science, The University of Tokyo, Tokyo 113-0033, Japan}
\altaffiltext{4}{Department of Physics, The University of Tokyo, 
Tokyo 113-0033, Japan}

\email{Electronic address: sho\_u@phys.se.tmu.ac.jp}
%%%%%%%%%%%%%%%%%%%%%%%%%%%%%%%%%%%%%%%%%%%%%%%%%%%%%%%%%%%%%%%%%%%%%%
\begin{abstract}
We visually inspected the light curves of 
7557 {\it Kepler} Objects of Interest (KOIs)
to search for single transit events (STEs) 
possibly due to long-period giant planets.
We identified 28 STEs in 24 KOIs, among which 14 events 
are newly reported in this paper.
We estimate the radius and orbital period of the objects causing STEs
by fitting the STE light curves simultaneously with the transits of the 
other planets in the system or with the prior information 
on the host star density.
As a result, we found that STEs in seven of those systems are
consistent with Neptune- to Jupiter-sized objects of orbital periods
ranging from a few to $\sim$ $20\,\mathrm{yr}$.
We also estimate that $\gtrsim20\%$ of the compact 
multi-transiting systems host cool giant planets 
with periods $\gtrsim 3\,\mathrm{yr}$ on the basis of
their occurrence in the KOIs with multiple candidates,
assuming the small mutual inclination between inner and outer planetary orbits.
\end{abstract}

\section{Introduction}\label{sec:intro}
About twenty years have passed since the first discovery of exoplanets 
using the radial velocity method \citep{1995Natur.378..355M}. 
Now the radial velocity surveys reach long-period exoplanets 
around and beyond the snow line, including Jupiter-analogues 
around sun-like stars \citep[e.g.][]{2012A&A...545A..55B,2013A&A...551A..90M,2015A&A...581A..34B, 2015arXiv151200417R}. 
How do we further characterize those cool gas and ice giants? 
While direct imaging is a promising approach to characterize them 
in near future \citep[e.g.][]{2010lyot.confE..22H,2014IAUS..299...66S},
transiting long-period giant planets (LPGs), 
on which the present paper focuses,
are also important for probing the planetary system architecture 
beyond the snow line. 
Indeed, detailed information on the system architecture, including 
the statistical properties of resonance and the mutual orbital inclination,
has already been obtained for the compact multi-transiting systems 
(orbital periods $\lesssim1\,\mathrm{yr}$) discovered 
by the {\it Kepler} spacecraft 
\citep[e.g.][]{2011ApJS..197....8L, 2014ApJ...790..146F, 
2015ARA&A..53..409W}. 
Transiting LPGs will also provide the opportunity to characterize 
both the interior structure and atmospheric compositions 
of the cool giant planets with transmission spectroscopy,
as already demonstrated for the solar-system planets.
For example, the observed transmission spectra of Jupiter 
\citep{2015ApJ...801L...8M} and Saturn \citep{2015arXiv151003430D} 
exhibit clear features of atmospheric molecules such as methane. 

Despite their importance, it is extremely challenging to find the
transiting LPGs at all.
Since the transit probability of LPGs is quite low ($\sim 0.1\%$),
it is hopeless to search for them in the sample of LPGs
characterized with radial velocities (RVs).
Even utilizing the space-mission data as obtained by {\it Kepler},
transiting LPGs can hardly be detected with the usual periodicity analysis
because their orbital periods are typically beyond the mission lifetime.
Nevertheless, they can still be detected through the 
single transit events (STEs), which occur only once in the $4$-year
observational span and thus may have been missed 
by the pipeline.\footnote{Note that the independent work by \citet{2015arXiv151202559W}, 
published during the preparation of this manuscript,
is based on a similar motivation to ours. The STEs in five systems we identified (KOI-4307, KOI-3349, KOI-847, KOI-1168, and KOI-3145) have also been reported in their paper.}

In this paper, we perform a uniform search for the STEs in the {\it Kepler} data
by visual inspection and report on the discovery of seven candidates
of transiting LPGs.
We focus on the targets with already known transit signals
(i.e. {\it Kepler} Objects of Interest, KOIs) mainly for the two reasons.
First, it is more likely to find transiting LPGs for those systems
because the orbital planes of LPGs are presumably aligned
with those of the inner planets, at least, to some extent.
Second, we can estimate the orbital periods of those transiting LPGs
even from the single transit, if found, 
by fitting their light curves simultaneously with those of the inner planets.
We will demonstrate that such an estimate is indeed possible
and also useful for discussing the system architecture.

The rest of the paper is organized as follows. 
\S \ref{sec:ste} presents the methods for finding the STEs 
and the analysis for estimating their geometric parameters,
especially the orbital period. 
\S \ref{sec:individual} describes the features of the individual STEs 
in more detail,
and classifies them based on the likelihood to be genuine planets.
Implications of our finding for the statistical property of LPGs
are also briefly discussed in \S \ref{sec:discussion}.

\section{Identification of Single Transit Events
and Orbital Period Estimation}\label{sec:ste}

Let us first summarize the STEs we identified before the detailed description of the analysis. Table \ref{tab:list} reports 28 STEs we identified in 24 KOIs.
We analyzed 16 STE light curves in 14 of those systems
that are not the clear false positives
%dispositioned as possible false positives by the {\it Kepler} team, 
and estimated the parameters including the orbital period
and radius ratio (Table \ref{tab:list_fit}).
As a result, we found seven systems exhibiting STEs consistent
with the planetary transit (\S 3 in detail); their orbital architectures 
are illustrated in Figure \ref{fig:sum}.
In this section, we describe the details of the STE identification 
(\S 2.1) and light-curve analyses (\S 2.2).

\subsection{Identification of Single Transiting Events (STEs)}
%We analyzed the long cadence data of the Pre-search Data Conditioning component of the {\it Kepler} pipeline (PDC-SAP).  
%The PDC-SAP data include the corrections of the instrumental signal and systematic trends in the target light curves. We analyzed all of long cadence data of Kepler Objects of Interests (KOIs) available on the NASA Exoplanet Archive at the time of June 4-th, 2015, which contains 7557 systems. 
%Long period modulations for 0.1-10 days in the light curves likely originating from the stellar activity were detrended. 
%We masked all transits (and secondary eclipses, if exist) 
%of known KOI candidates and searched for STEs by visual inspection.
We analyzed the long cadence fluxes of the 
Pre-search Data Conditioning component of the {\it Kepler} pipeline 
(PDCSAP) of 7557 KOIs, which were
available on the NASA Exoplanet Archive at the time of June 4th, 2015.
We searched for STEs by visual inspection of all those PDCSAP fluxes
and identified 28 STEs in 24 KOIs as shown in Table \ref{tab:list}.
Here the fading events that are not observed in the SAP data
but only seen in the PDCSAP data are excluded
because we find that the correction by the PDC pipeline 
sometimes leads to artificial dips in the light curves.\footnote{For instance, PDCSAP data of KOI-6469 (KIC 4912589) 
exhibits a dip at $\mathrm{BJD}-2454833 = 613.5$,
which does not exist in the SAP data.}
Although it is admittedly difficult to quantify the detection limit of our 
visual inspection, we believe that transits deeper than $\sim 0.1\%$ 
and lasting longer than $5\,\mathchar`-50$ hours have been detected.

Among the 28 STEs in Table \ref{tab:list}, 
14 have never been reported in the literature; 
they are marked with ``new" in the parentheses. 
KOI numbers designated by the {\it Kepler} team are listed for the 
other seven events.
When more than one STEs are found in one system, they are reported
separately (two in KOI-847, KOI-1168, and KOI-6378; three in KOI-1032).

\subsection{Geometric Parameters of STE Candidates}
To further characterize the planet candidates causing STEs 
(hereafter ``STE candidates"), we fit the STE light curves assuming that
STE candidates are not due to contamination 
but orbiting the KOIs for which we found STEs on circular orbits.
As discussed below, this assumption allows us to
estimate orbital periods of the STE candidates  
even from only one transit.
Here we exclude the systems
\red{designated as false positives in the KOI catalog
and KOI-1032 exhibiting signatures of the CCD cross talk\footnote{
\red{This system is classified as a possible false positive
on Kepler Community Follow-up Observing Program (CFOP) webpage as well.}}
}
%Here we excluded the systems designated as possible false positives 
%by the {\it Kepler} team\footnote{
%Based on Kepler Community Follow-up Observing Program (CFOP) webpage.}
because the above assumption is less sound for them.
We also examined the target pixel files of the 
\red{remaining targets visually}
and excluded the STE of KOI-3145, \red{whose depths are 
different in neighboring pixels and thus likely to be due to
contamination from a nearby star \citep[see also][]{2015arXiv151202559W}.
}
These criteria leave us with 16 STEs in 14 systems,
which are listed in Table \ref{tab:list_fit} along with their estimated parameters.

\subsubsection{Principle}
While we use the Markov-Chain Monte Carlo (MCMC) method
to determine the system parameters, 
here we analytically show how the orbital period of the STE candidate 
is derived with the information on the mean stellar density either from 
the light curves of inner transiting planets with known orbital 
periods or from the follow-up observations of the host star.

Assuming a circular orbit, the total and full transit durations 
(denoted by $t_T$ and $t_F$, respecitvely) 
are given by \citep[e.g.][]{2010arXiv1001.2010W} 
\begin{eqnarray}
\label{eq:totald}
t_T = \frac{P}{\pi} \sin^{-1} \left(\frac{R_*}{a} \sqrt{\frac{(1+k)^2 - b^2}{\sin{i}}}\right),\\
%\sin{ \left( \frac{\pi t_T}{P} \right)}  = \frac{R_*}{a} \sqrt{\frac{(1+k)^2 - b^2}{\sin{i}}}, \\
\label{eq:totalf}
t_F = \frac{P}{\pi} \sin^{-1} \left(\frac{R_*}{a} \sqrt{\frac{(1-k)^2 - b^2}{\sin{i}}}\right),
%\sin{ \left( \frac{\pi t_F}{P} \right)}  = \frac{R_*}{a} \sqrt{\frac{(1-k)^2 - b^2}{\sin{i}}}, 
\end{eqnarray}
where $k \equiv R_p/R_\ast$ indicates 
the ratio of the planetary radius to the stellar radius, 
$i$ is the orbital inclination, and $b$ is the impact parameter. 
%On the approximation that $\sin{(\pi t_T/P)} \sim \pi t_T/P$ and $\sin{(\pi t_F/P)} \sim \pi t_F/P$, 
Neglecting the terms of $\mathcal{O}((R_\ast/a)^3)$ and higher, 
we obtain
\begin{eqnarray}
\label{eq:ra}
\frac{R_\ast}{a} = \frac{\pi}{2 \sqrt{k}} \frac{\sqrt{t_T^2 - t_F^2}}{P}.
\end{eqnarray}
Combined wtih Kepler's third law, Equation (\ref{eq:ra}) yields
the orbital period of an STE candidate 
in terms of its transit shape and the mean stellar density $\rho_*$ \citep{2003ApJ...585.1038S}:
\begin{eqnarray}
\label{eq:pk}
P = \frac{\pi G}{32} \frac{M_\ast}{R_\ast^3} \left( \frac{t_T^2 -t_F^2}{k} \right)^{\frac{3}{2}}
\simeq \frac{\pi^2 G}{3} \rho_* \left( \frac{T\tau}{k} \right)^{3/2},
\end{eqnarray}
where $M_\ast$ and $R_\ast$ are the stellar mass and radius.
We also defined $T \equiv \frac{1}{2}(t_T + t_F)$ and $\tau \equiv \frac{1}{2}(t_T - t_F)$ and assumed $\tau \ll T$.

If the inner transiting planet(s) is known in the system, 
Equation (\ref{eq:pk}) can be used to constrain $\rho_*$ 
from its transit shape because $P$ is already determined 
for the planet(s).
In this case, the orbital period of the STE candidate $P_s$ is given by
\begin{equation}
\label{eq:ps_analytic}
P_s = \left( \frac{k_i T_s \tau_s}{k_s T_i \tau_i} \right)^{\frac{3}{2}} P_i,
\end{equation}
where the subscripts $s$ and $i$ denote the quantities 
for the STE candidate and inner planet, respectively.
This is the case for the 10 systems except for KOI-1421, 1208, 1174,
and 1096 in Table \ref{tab:list_fit}.
We checked that the analytic estimate in Equation (\ref{eq:ps_analytic})
is indeed consistent with the MCMC results for these systems.
Even if inner transiting planets are not known, 
the prior knowledge on $\rho_*$ from the color photometry can 
also be used to constrain $P_s$, although it may be less reliable
than the dynamical value as obtained in the previous case;
this method is adopted for KOI-1421, 1208, 1174, and 1096.

\subsubsection{MCMC Fit to the Observed STE Light Curves}
For the 10 systems except for KOI-1421, 1208, 1174, and 1096,
we fit the STE light curves simultaneously with the phase-folded
transit light curves of the other planet candidates in the system.
We basically analyzed the long-cadence PDCSAP fluxes
except for the inner transits of KOI-671 and KOI-435,
for which short-cadence fluxes were used.
We used {\tt PyTransit} package \citep{2015MNRAS.450.3233P}
to generate the transit light curve based on \citet{2002ApJ...580L.171M}
model for the quadratic limb-darkening law. 
The effect of binning was taken into account by supersampling 
for the long-cadence data sampled at 30 minutes.
Constraints on the parameters were obtained by the MCMC sampling
using {\tt emcee} package \citep{2013PASP..125..306F} following 
the standard $\chi^2$ minimization using the Levenberg-Marquardt method \citep[{\tt mpfit} by][]{2009ASPC..411..251M}.
The likelihood for the MCMC sampling $\mathcal{L}$ was computed as 
$\mathcal{L}\propto\exp(-\chi^2/2)$, where $\chi^2$ is a sum of the 
standard chi-squared for each planet's transit,
and we adopt non-informative priors unless otherwise noted.

Transits of the inner planets were processed as follows. 
We first detrended the light curves from each quarter
using the second-order spline interpolation, 
after masking the known transits based on the 
ephemeris and duration in the KOI catalog. 
Baseline fluxes during the transit were determined by the linear 
interpolation between the two ends of the masked region.
The detrended transits were folded at the orbital period 
given in the KOI catalog and averaged into one-minute bins.
The value and error of the binned flux are given by the mean
and its standard deviation of the values in each bin.
The smoothing parameter of the spline was chosen so that
the depth of the phase-folded transit be consistent with the catalog value.
The STE light curves were deterened in a similar manner
except that the smoothing parameter was chosen to be 
$0.1\,\mathrm{days}$
and that the endpoints of the linear interpolation were adjusted
so that the resulting detrend light curve be not too asymmetric. 

The fitting parameters in this case are
the sum of two coefficients for the quadratic limb-darkening law 
$u_1+u_2$ and $\rho_\ast$ as the parameters common to all the planets
in the same system;
time of transit center $T_c$, $\cos i$, $k$, and normalization of the flux $c$
for each of the inner transiting planets;
and $T_c$, $\cos i$, $k$, $c$, and orbital period $P_s$ for STE candidates.
The difference of the limb-darkening coefficients $u_1-u_2$ were
fixed to be the linearly interpolated values 
based on \citet{2010A&A...510A..21S},
and the orbital periods of the inner planets were fixed at the values
given in the KOI catalog.
In MCMC fitting, we adopt the prior distributions uniform in
$u_1+u_2$, $\log \rho_\ast$, $T_c$, $\cos i$, $\log k$, $c$, and $\log P_s$.
We assumed circular orbits for both inner candidates and STE candidates.

For KOI-1421, 1208, 1174, and 1096, 
we fit the STE light curve alone, imposing the 
Gaussian prior on $\rho_\ast$ based on the value at CFOP.
The fitting parameters in this case are the two stellar parameters
$u_1+u_2$ and $\rho_\ast$,
and $T_c$, $\cos i$, $k$, $c$, and $P_s$ of the STE candidate.

\red{
In the analyses above, we neglect the effect of possible stellar multiplicity.
If the target star has an unidentified companion star, for instance,
the planetary radii can be underestimated due to the dilution.
We note, however, that this possibility is essentially ruled out for most 
of our main candidates discussed in Section 3.1.
}

\begin{figure}[htbp]
\begin{center}
\includegraphics[width=\linewidth,clip]{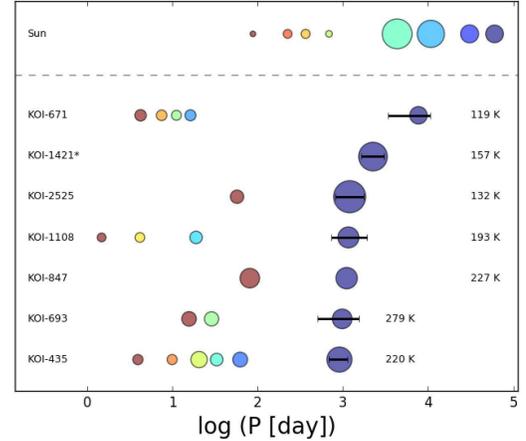}
\caption{\label{fig:sum} Architectures of the KOI systems 
for which we identified STEs consistent with 
transiting LPGs (listed in the upper part of Table \ref{tab:list_fit}).
The size of the circles are proportional to the estimated planet radius. 
The STE candidates are the rightmost circles in each row, 
illustrated with error bars for the estimated period and its
radiative equilibrium temperature.
Note that the orbital period determined from 
the interval of the two STEs is adopted for KOI-847.
Our solar system is shown at the top for reference. 
The orbital period of the STE candidate in KOI-1421 
(marked with an asterisk) is based on the stellar density 
provided by CFOP and may be less reliable than the others.
}
\end{center}
\end{figure}

\begin{figure*}[htbp]
\begin{center}
\includegraphics[width=0.85\linewidth]{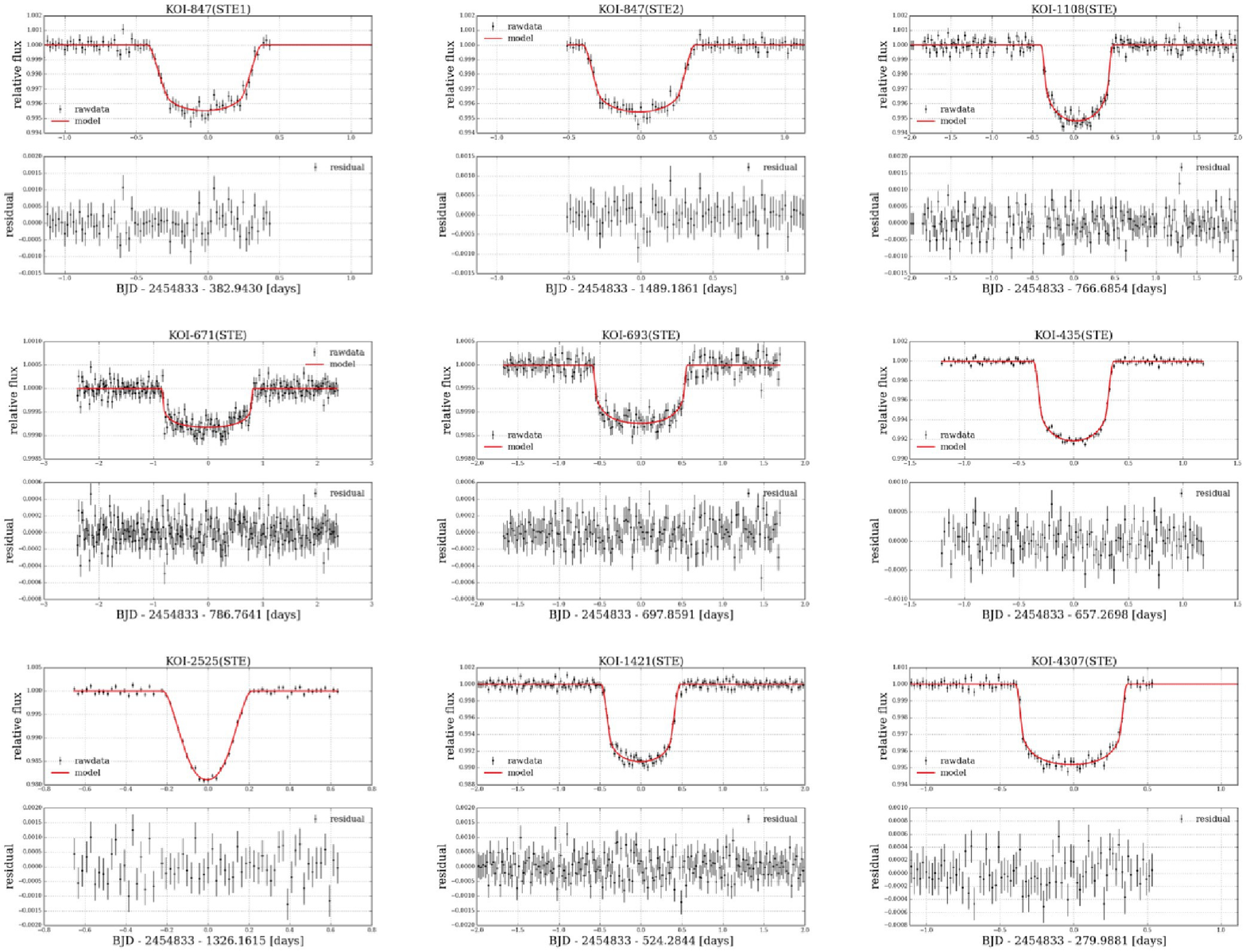}
\includegraphics[width=0.85\linewidth]{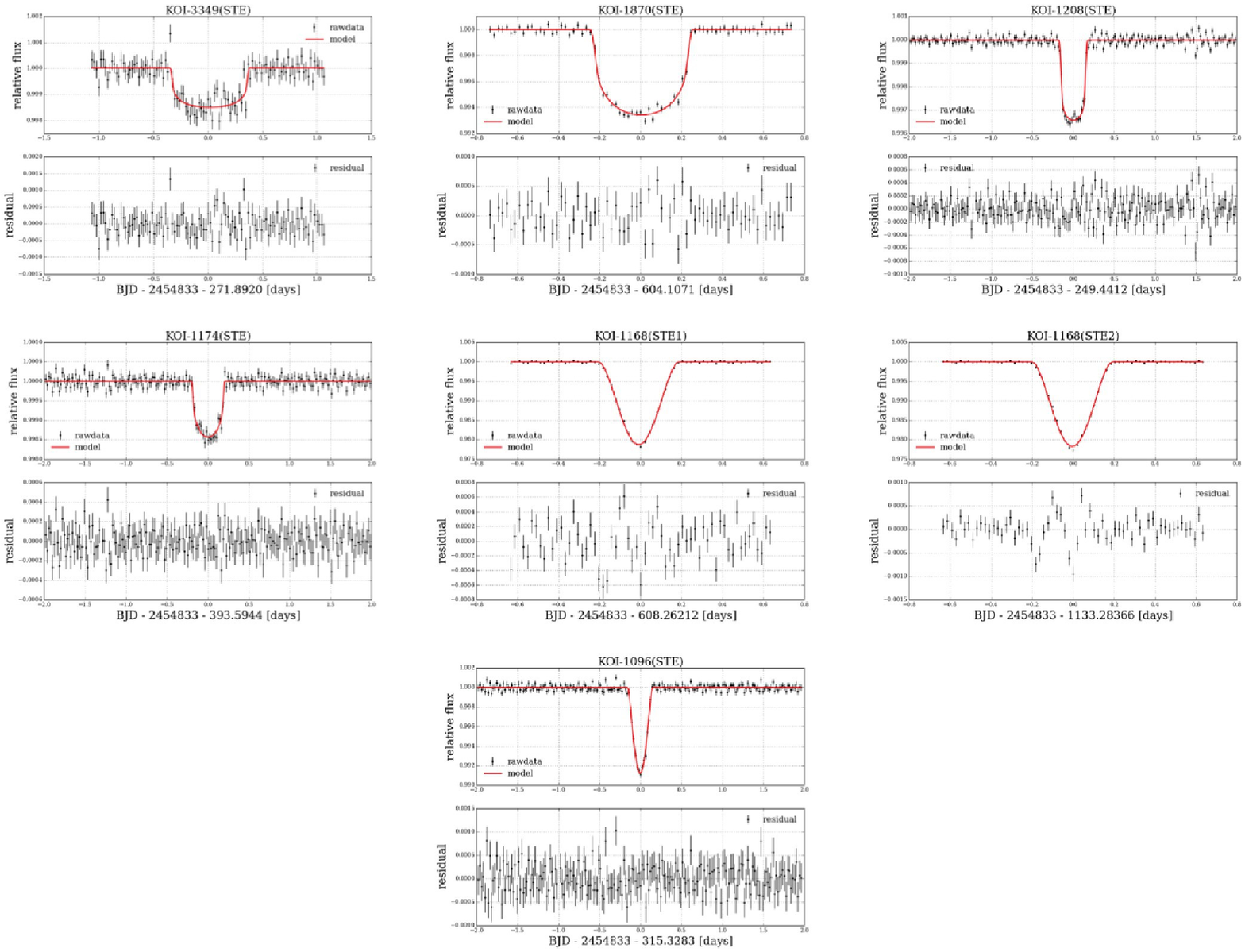}
\caption{Light curves of the 16 STEs we analyzed. The solid red lines 
show the best-fit model obtained by $\chi^2$ minimization. \label{fig:fit_all}}
\end{center}
\end{figure*}

\begin{figure*}[htbp]
\begin{center}
\includegraphics[width=0.75\textwidth]{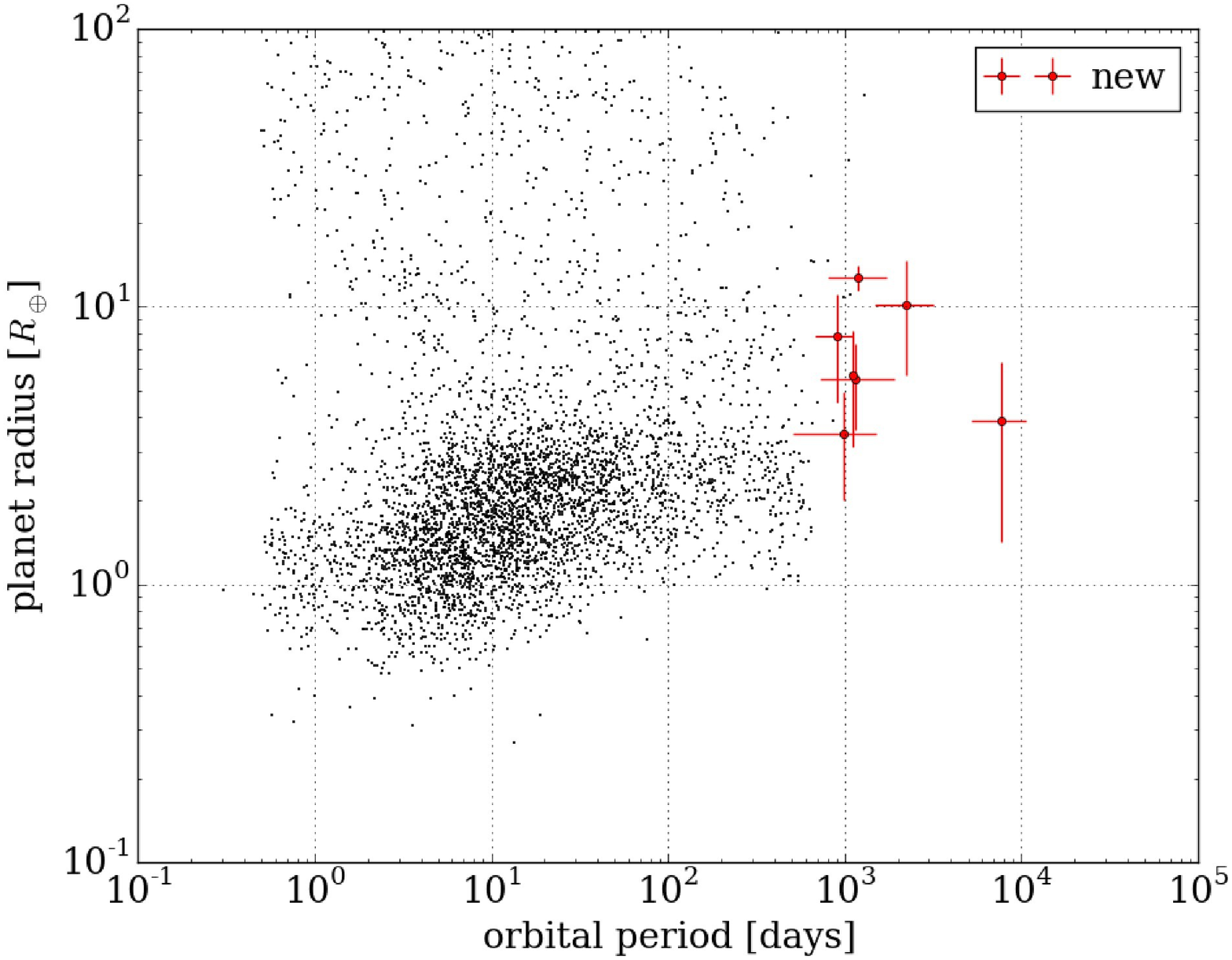}
\caption{Properties of the seven STE candidates in Figure \ref{fig:sum} 
on the period-radius plane (red dots with error bars). 
All known KOI candidates are also shown by small black dots. \label{fig:p}}
\end{center}
\end{figure*}

The best-fit models and constraints on the parameters of STE candidates
are summarized in Figure \ref{fig:fit_all} and Table \ref{tab:list_fit}.
All the transits of the inner candidates simultaneously fitted with STEs 
are listed in Appendix.
On the basis of the inferred orbital period and radius of STE candidates,
we found that STEs in seven systems are consistent with the planetary transit,
as will be detailed in \S \ref{sec:individual}.
Architectures of these seven systems are illustrated in Figure \ref{fig:sum}.
We also plotted each STE candidate on the $R_p\,\mathchar`-P$ plane 
with all known KOI candidates in Figure \ref{fig:p}. 
The figure shows that STE candidates we found are all likely to be
gas/ice giants beyond the snow line. 

%%%%%%%%%%%%%%%%%%%%%%%%%%%%%%%%%%%%%%%
%%%%%%%%%%%%%%%%%%%%%%%%%%%%%%%%%%%%%%%

%%%%%%%%%%%%%%%%%%%%%%%%%%%%%%%%%%%%%%%
%%%%%%%%%%%%%%%%%%%%%%%%%%%%%%%%%%%%%%%

%%%%%%%%%%%%%%%%%%%%%%%%%%%%%%%%%%%%%%%
%%%%%%%%%%%%%%%%%%%%%%%%%%%%%%%%%%%%%%%

\section{Classification of STE Candidates and 
Description of Individual Systems}\label{sec:individual}
On the basis of the fitting results,
we classify the STE candidates in Table \ref{tab:list_fit} into three categories:
\begin{enumerate}
\item candidates consistent with transiting LPGs,
\item candidates exhibiting anomalous orbital periods, which
we call ``misfit" singles, and
\item candidates likely to be eclipsing binaries (EBs). 
\end{enumerate}
In the following, we describe the details of the classification
and comment on the properties of individual systems.

\subsection{STE Candidates Consistent with the Transiting LPGs}
Seven systems illustrated in Figure \ref{fig:sum} exhibit STEs 
consistent with the planetary transit.
Namely, 
(1) their transit light curves are consistently explained with those 
of the inner transiting planets or stellar density $\rho_\ast$ in the KIC catalog,
(2) their inferred radii are less than about the Jupiter radius, and
(3) their inferred orbital periods are consistent with the absence of 
other transits in the exiting {\it Kepler} data 
(i.e. inferred $P_s$ is longer than $P_{\rm min, Kepler}$ in Table \ref{tab:list}).
\red{It is worth noting that most of the candidates in this category 
except for KOI-1421 are in multi-planetary systems 
and thus likely to be genuine planets in the same systems
\citep{2014ApJ...784...44L}.}

\subsubsection*{KOI-847 (KIC 6191521)}
A Neptune-sized planet candidate with the period of 80.9 days (KOI-847.01) 
is already known in this system. 
We found additional two transit at 
\red{$T_c (\mathrm{BJD} - 2454833) = 382.9428$ and $1489.1858$}. \footnote{
These events are reported independently by \citet{2015arXiv151202559W} as well.}
If these two transits originate from the same object, 
its orbital period $P_s$ is $1106.243\pm0.007\,\mathrm{days}$. 
We individually fit each of the two events with the inner transit light curves
and found the periods consistent with the above interval,
\red{$P_s =930^{+430}_{-380}\,\mathrm{days}$
and $840^{+380}_{-300}\,\mathrm{days}$, respectively.}
In addition, the other transit parameters were also consistent 
with each other. The two STEs are therefore likely to be attributed to
the same Neptune-sized planet with $P_s=1106.243\pm0.007\,\mathrm{days}$.

% 3145 removed
%\subsubsection*{KOI-3145 (KIC 1717722)}
%Two super-Earth candidates with the periods of 0.977 days and 4.53 days (KOI-3145.01 and KOI-3145.02) are already known in this system. 
%We found one transiting event at $T_c (\mathrm{BJD} - 2454833) = 1439.1973$ consistent with a transit of a Jupiter-sized object with $R_p=11.7\pm1.2 R_\oplus$.
%The inferred orbital period $P_s = 6900^{+2500}_{-1900}\,\mathrm{days}$ is the longest among the STE candidates we found and is between those of Jupiter and Saturn. The radiative equilibrium temperature is $\sim$ 80 K. This planet candidate probably formed beyond the snow line will be the promising target to characterize the cool gas giants in future.

%Another noticeable feature of this STE is the systematic 
%residuals found during the egress, which may be explained
%by a ring structure similar to that of Saturn, an exomoon, or the red noise. 
%However, detailed modeling of the feature is beyond the scope of this paper.

\subsection*{KOI-671 (Kepler-208, KIC 7040629)}
KOI-671 is known as a compact multi-transiting system
hosting four transiting planet candidates inside the orbit of Mercury
(KOI-671.01--04). 
We found an STE at \red{$T_c (\mathrm{BJD} - 2454833) = 786.7641$}
consistent with a Neptune-sized planet of
$R_p=3.9\pm2.5\,R_\oplus$ and $P_s=7700^{+2900}_{-2500}\,\mathrm{days}$.
This $P_s$ is the longest among our STE candidates
we found and is between those of Jupiter and Saturn.

\subsubsection*{KOI-2525 (KIC 5942949)}
This system has one super-Earth candidate (KOI-2525.01). 
We found an STE at \red{$T_c (\mathrm{BJD} - 2454833) = 1326.1614\,\mathrm{days}$}, for which we found 
\red{$R_p=12.7\pm1.3\,R_\oplus$} and 
\red{$P=1200^{+540}_{-390}\,\mathrm{days}$}
from the MCMC analysis.

\subsubsection*{KOI-1108 (KIC 3218908)}
KOI-1108 has a compact multi-transiting system 
with three super-Earth candidates within $P=18\,\mathrm{days}$
(KOI-1108.01--03). 
We found an STE by a Neptune-sized object 
\red{
($R_p = 5.5 \pm 1.9\,R_\oplus$) with $P=1160^{+760}_{-430}\,\mathrm{days}$.}

\subsubsection*{KOI-693 (Kepler-214, KIC 8738735)}
This system harbors two confirmed super Earths Kepler-214b and c. 
We found one STE corresponding to a Neptune-sized object with
$R_p = 3.5\pm1.5\,R_\oplus$ and $P=980^{+520}_{-470}\,\mathrm{days}$.

\subsubsection*{KOI-435 (Kepler-154, KIC 11709124)}
The system hosts two confirmed super-Earths/sub-Neptunes
(Kepler-154a and b) and three planet candidates of similar radii.
We found an STE due to a Saturn-sized object at 
$T_c (\mathrm{BJD} - 2454833) = 657.2698$, 
which is listed as KOI-435.02 in the KOI catalog. 
Our estimated radius and period are consistent with 
those in the catalog. 
The orbital period is best constrained for this STE candidate 
(except for KOI-847 exhibiting two STEs)
due to the presence of multiple inner transiting planets 
and high signal-to-noise ratio. 

\subsubsection*{KOI-1421 (KIC 11342550)}
The STE we found is also listed in the KOI catalog as KOI-1421.01.
Since this is the only transit signal known for the system,
we fit the single STE light curve with a Gaussian prior 
on the mean stellar density $\rho_\ast = 1.403 \pm 0.3668\,\mathrm{g/cc}$,
which is based on the CFOP value.
As a result, we obtained $P=2230^{+960}_{-740}\,\mathrm{days}$ 
and $R_p=10.2\pm4.5\,R_\oplus$.
The orbital period is consistent with the value given in the KOI catalog.
\red{As mentioned at the beginning of this subsection,
this candidate has a higher false positive probability than the others
because this system has no inner companion.
However, we ruled out the companion identified 
at $11^{\prime\prime}$ away as a source of the fading event
from inspection of the target pixel file.} 

%\subsection{The STE with no multiple transit KOI candidates (S-singles)}
%In the category we describe in this section, we include the STE around the KOI "FALSE POSITIVE"s or the STE those the KOI catalog already identified.
%For the STE, the orbital period is estimated by assuming the stellar density provided by CFOP. In these cases, we feel less confident about the estimated period because of the possibility of contaminations and systematic uncertainty of the stellar density estimates.  
%\subsubsection*{KOI-99 (KIC 8505215)}
%This system has one STE. Using the stellar density $\rho_\ast =3.139 \pm 0.2093$ given by CFOP, we derived $P=4500^{+4500}_{-1600}$ day and $R_p = 3.3 R_\oplus$. This target is the second brightest among the STE systems we found. This event is listed in the KOI catalog. The KOI catalog provides $P=2190$ day, which is slightly smaller than the value we estimated.   
%\subsection*{KOI-3475 (KIC 6145201), KOI-7194 (KIC 9581498), KOI-154 (KIC 9970525)}
%For these candidates, we do not estimate the orbital period because the stellar density is not available. 

\subsection{``Misfit" Singles: 
KOI-4307, KOI-3349, KOI-1870, KOI-1208, KOI-1174, and KOI-1096
--- Eccentric Planets of False Positives?}
For the STEs in these systems, the orbital periods inferred from our fitting
are shorter than $P_{\rm min, Kepler}$ in Table \ref{tab:list},
which is the minimum orbital period required for the STE candidate
to be consistent with the absence of other transit signals in the
existing {\it Kepler} data.
While the discrepancy may imply that they are false positives,
it is also possible that the orbits of these STE candidates are 
eccentric, as shown below.
For this reason, we still consider these ``misfit" singles as planet candidates,
though less promising than those listed in the previous subsection.

For simplicity, we consider the case where only the STE candidates have 
non-zero eccentricities, while the inner transiting objects are all
on circular orbits.
For eccentric orbits, the right-hand sides of $t_T$ and $t_F$ 
in Equations (\ref{eq:totald}) and (\ref{eq:totalf}) are 
multiplied by $\sqrt{1-e^2}/(1+e\sin\omega)$, where
$e$ is the eccentricity and $\omega$ is the argument of periastron
(measured from the sky plane) of the STE candidate \citep[e.g. equation (16) of][]{2010arXiv1001.2010W}.
Thus, the orbital period of the STE candidate with non-zero eccentricity
differs from the circular case by the factor of 
\begin{equation}
	\alpha_{\rm ecc} = \left( \frac{1+e\sin\omega}{\sqrt{1-e^2}} \right)^3
	\leq \left(\frac{1+e}{1-e}\right)^{3/2}
	\label{eq:p_ratio}
\end{equation}
assuming the same $\rho_\ast$.\footnote{While the joint fit including STE candidate's eccentricity
may change $\rho_\ast$ as well, we neglect such rather small effects
for the rough estimate here.}
Since $\alpha_{\rm ecc}$ is larger than the ratio of $P_{\rm min, Kepler}$
to $P_s$ obtained from the circular fit (Table \ref{tab:list_fit}), i.e.,
$\alpha_{\rm ecc}>P_{\rm min, Kepler}/P_s$,
Equation (\ref{eq:p_ratio}) yields the minimum value of eccentricity 
required to explain the observation:
\begin{equation}
	e_{\rm min} 
	= \frac{(P_{\rm min, Kepler}/P_s)^{2/3}-1}{(P_{\rm min, Kepler}/P_s)^{2/3}+1}.
\end{equation}
The values of $e_{\rm min}$ and $P_{\rm min, Kepler}$
for each of the ``misfit" candidates are
listed in Table \ref{tab:list_misfit}, 
along with their values computed for the most conservative case 
(see the note in the table).

If (some of) these candidates are confirmed to be 
eccentric LPGs by follow-up observations,
they can be interesting targets to understand the origin of 
hot Jupiters, as in the case of HD 80606 \citep{2003ApJ...589..605W}.
In this regard, the most promising target is KOI-1208
with the {\it Kepler} magnitude of 13.6 (Table \ref{tab:list}), 
for which a possible nearby companion has been detected 
and the eccentricity is estimated to be $\gtrsim 0.8$.

\subsection{Candidate Likely to be an EB: KOI-1168}
%, KOI-1032, KOI-1096, KOI-2824, KOI-6378}
The inferred radius of this STE candidate suggests that
it is a stellar object rather than a planet.
The good agreement between the parameters of the two STEs
strongly implies that they are due to the same object.
If this is indeed the case, 
this objects has $P_s=525.0216\pm0.0006\,\mathrm{days}$.
This system is also discussed in \citet{2015arXiv151202559W},
who also classified it as a likely false positive.

%%%%%%%%%%%%%%%%%%%%%%%%%%%%%%%%%%%%%%%
\section{Discussion}\label{sec:discussion}
\subsection{Gap-like Structure}
In Figure \ref{fig:sum}, the systems with multiple inner planets
(KOI-671, 1108, 693, and 435) exhibit a gap-like signature
between the inner and STE candidates,
as seen between the terrestrial and giant planets in the solar system (top row).
While we cannot completely exclude the existence of such a gap in our sample,
we suspect that it is due to the decrease in the transit probability 
with increasing orbital periods. Because of the low transit probability of a distant planet, transiting objects on wide orbits (say $P\gtrsim10^3\,\mathrm{days}$) tend to be rare. This means that the system with one long-period transiting object detected, as in our sample, is unlikely to have another transiting object close to 
the detected one and exhibits a gap around the detected transiting object.

\begin{figure}[htbp]
\begin{center}
\includegraphics[width=\linewidth,clip]{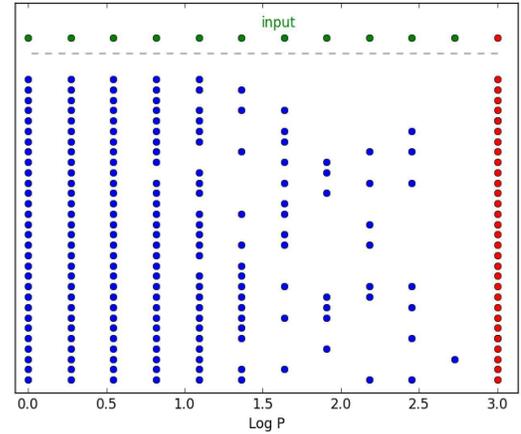}
\caption{Simulated system architectures detected 
by transit observations.
The top row shows the architecture of the input planetary system.
The other rows plot the planets detected with simulated transit observations
(blue circles) for the systems whose outermost planet (red circles) transits.
 \label{fig:sim}}
\end{center}
\end{figure}

To demonstrate the above geometric effect, 
we performed a simple simulation of the transit detection 
of the multi-planetary system. 
We put 12 planets at a constant log-interval as in the first row 
of Figure \ref{fig:sim}.
The semi-major axis over the host star radius, $a/R_\star$,
of the outermost planet was chosen to be $420$,
which corresponds to $P=10^3\,\mathrm{days}$
or $a=1.96\,\mathrm{AU}$ in the solar system.
We simulated the transit observation of this system from many different
directions uniformly distributed in $\cos i$,
sampling the mutual inclinations of the inner planets with respect to the
outermost one from the Rayleigh distribution with $\sigma=1\fdg8$
\citep{2014ApJ...790..146F} in each run.
From the resulting sample of transiting systems, 
we chose the systems where the transit of the outermost planet was detected
and plotted their apparent architecture below the horizontal dashed line 
in Figure \ref{fig:sim}.
One can actually see the gap-like structure between the outermost planets 
(red circles) and the inner planets (blue circles) in many cases.

\subsection{Occurrence Rate of the LPG in Compact Multi-transiting Systems}

As summarized in Table \ref{tab:occurrence_lpgs}, 
the sample of 7557 KOIs we surveyed includes 695 systems with
more than one inner transiting planet candidates, among which
we detected four LPG candidates (see also Figure \ref{fig:sum}).
These numbers can be used to estimate the occurrence rate of LPGs
in compact multi-transiting systems
on the premise that LPG orbits are likely to be well aligned 
with those of the inner multiple planets.
The premise is based on the following argument.
The planets in a compact multi-transiting system
presumably have well-aligned orbits \citep{2014ApJ...790..146F},
indicating that their orbital planes trace the original protoplanetary disk.
Since an LPG in the same system formed in the same disk,
its orbit is likely to be aligned with the inner ones as well.

The expected number of transiting LPGs, $n_{\rm tLPG}$, 
in $N_{\rm cmulti}$ compact multi-transiting systems is given by
\begin{equation}
	\label{eq:occurrence}
	n_{\rm tLPG} 
	\simeq \frac{T_{\rm obs}}{P_{\rm LPG}}\,
	p(\mathrm{tra}|\mathrm{LPG}, \mathrm{cmulti})\,
	\overline{n}(\mathrm{LPG}|\mathrm{cmulti})\,
	N_{\rm cmulti},
\end{equation}
where $\overline{n}(\mathrm{LPG}|\mathrm{cmulti})$ 
is the average number of LPGs per system and
$p(\mathrm{tra}|\mathrm{LPG}, \mathrm{cmulti})$ 
is the transit probability of a given LPG,
both under the existence of the inner compact multi-transiting system.
The factor $T_{\rm obs}/P_{\rm LPG}$, where $T_{\rm obs}$ is the
observing duration of {\it Kepler}
and $P_{\rm LPG}$ is the orbital period of a given LPG,
takes into account the probability that the single transit of the LPG
falls into the mission life time of {\it Kepler}.
Since $p(\mathrm{tra}|\mathrm{LPG}, \mathrm{cmulti})$
depends on the mutual inclination of the LPG and the inner planets, 
the mutual inclination and occurrence rate $\overline{n}$ 
are usually degenerate \citep{2012AJ....143...94T}.

As far as the mutual inclination between the LPG and inner-planet orbits
is as small as $\sim R_\star/a_{\rm in}$,
we approximately have
$p(\mathrm{tra}|\mathrm{LPG}, \mathrm{cmulti}) = a_{\rm in}/a_{\rm LPG}$,
where $a_{\rm LPG}$ and $a_{\rm in}$ 
are the typical semi-major axes of the LPG and inner multi-transiting planets,
respectively \citep[see \S2.3 of][]{2010arXiv1006.3727R}.
Adopting $a_{\rm LPG}=2\,\mathrm{AU}$ 
(corresponding to $P\simeq10^3\,\mathrm{days}$)
and $a_{\rm in}=0.07\,\mathrm{AU}$ (median of KOI candidates in multi-transiting systems),
we obtain $p(\mathrm{tra}|\mathrm{LPG}, \mathrm{cmulti})=0.035$,
which yields
$\overline{n}(\mathrm{LPG}|\mathrm{cmulti}) \simeq 0.2$
for $n_{\rm tLPG} =4$, $N_{\rm cmulti}=695$, $T_{\rm obs}=4\,\mathrm{yr}$,
and $P_{\rm LPG}=2200\,\mathrm{days}$ (average of the seven systems
in Figure \ref{fig:sum}) in Equation (\ref{eq:occurrence}).

%We performed the following simulation to check that the above estimate is valid even in the presence of small but non-zero mutual inclination as estimated for the compact multi-transiting systems. We put six inner planets with $P=6\,\mathchar`-25\,\mathrm{days}$(corresponding to medians of the innermost/outermost candidates in multi-transiting KOI systems) and one LPG with $P=10^3\,\mathrm{days}$  and simulated the detection probability of the outer LPG given the detection of {\bf more than one} inner planets. The mutual inclinations of the inner six planets relative to the outer LPG were again sampled from the Rayleigh distribution with $\sigma=1\fdg8$. From this simulation, we found $p(\mathrm{transit}|\mathrm{LPG}, \mathrm{cmulti})=0.035$, which is close to the value computed for perfectly aligned orbits.

If the LPG has actually larger mutual inclination relative to the
inner planets than assumed here ($\gtrsim R_\star/a_{\rm in}\sim4^\circ$), 
the above estimate underpredicts the true occurrence rate.\footnote{If the inclination of the outer LPG is completely random, for instance,
$p(\mathrm{tra}|\mathrm{LPG}, \mathrm{cmulti})=R_\star/a_{\rm LPG}
\approx 0.002$ for $a_{\rm LPG}=2\,\mathrm{AU}$ and $R_\star=R_\odot$,
which results in $\overline{n}(\mathrm{LPG}|\mathrm{cmulti}) \simeq 3$.}
In addition, the above discussion assumes 
that the transiting LPG is $100\%$ detected
as long as it transits the host star during the {\it Kepler} observation.
For these reasons, we conclude that $\overline{n}(\mathrm{LPG}|\mathrm{cmulti}) \simeq 0.2$ estimated above is a rough lower limit,
and that about $20\%$ or more of the compact multi-transiting systems 
host LPGs with $P\gtrsim10^3\,\mathrm{days}$.

\subsection{Different Mutual Inclinations or Occurrence Rates
in Single- and Multi-transiting Systems?}
Table \ref{tab:occurrence_lpgs} also shows
that the fraction of transiting LPGs in KOIs with only one 
inner transiting candidate 
is smaller than the above multiple-candidate case 
by an order of magnitude.
While we suffer from the small statistics, the fact suggests that
the term $p(\mathrm{tra}|\mathrm{LPG}, \mathrm{cmulti})\,\overline{n}(\mathrm{LPG}|\mathrm{cmulti})/P_{\rm LPG}$ in Equation (\ref{eq:occurrence})
is smaller for (a part of) the single-candidate sample.
This means that either
(A) mutual inclination of the LPG relative to the inner planet 
may be larger, 
(B) the occurrence rate of the LPG may be smaller, or
(C) the typical orbital period of the LPG may be longer,
in the systems with only one inner transiting planet.
If (A) is actually the case, the result supports the scenario by
\citet{2014ApJ...796...47M}
that a population of highly inclined multi-planet systems contributes
the excess of single-transiting systems in the {\it Kepler} multiplicity statistics
\citep{2011ApJS..197....8L},
which is known as the ``{\it Kepler} dichotomy."

\acknowledgements
We are grateful to Masahiro Ikoma and Yasushi Suto for fruitful discussions, the {\it Kepler} team for their revolutionary data, and an anonymous referee for many helpful suggestion. H.K. is supported Grant-in-Aid for Young Scientists (B) from
Japan Society for Promotion of Science (JSPS), No.\,25800106.
K.M. is supported by JSPS Research Fellowships for Young Scientists 
(No.\,26-7182) and 
by the Leading Graduate Course for Frontiers of Mathematical Sciences 
and Physics (FMSP).
M.A. acknowledges the support by the
Advanced Leading Graduate Course for Photon Science (ALPS).

%\bibliography{ref} 

%\bibliography{/Users/k_masuda/Dropbox/ApJ_format/references_masuda}

%%%%%%%%%%%%%%%%%%%%%%%%%%%%%%%%%
%	Table
%	List of STEs
%%%%%%%%%%%%%%%%%%%%%%%%%%%%%%%%%
\begin{table*}[!tbh]
\begin{center}
\caption{List of the 28 Single Transit Events We Identified \label{tab:list}}
%\scalebox{0.9}{
\begin{tabular}{lccccccccc}
   \hline\hline
KepID  & KOI\tablenotemark{a} & architecture\tablenotemark{b} & Kepler mag & $T_c$ (KBJD) & depth (ppm)\tablenotemark{c} & $T_{\rm eff}$ (K) & $\log\,g$ (cgs) & $P_{\rm Kep, min}$ (days)\tablenotemark{d}\\
   \hline
8505215 & 99.01 & 1CS+1F & 13.0 & 140.0473 & 1874.2 & 4965 & 4.555 & 1450.9538 \\
9970525 & 154(new) & 1F+1S & 13.2 & 139.7277 & 1500 & 6504 & 4.355 & 1451.2732 \\
11709124 & 435.02& 5C+1CS & 14.5 & 657.2698 & 8709.6 & 5937 & 4.559 & 933.7415 \\ 
7040629 & 671(new) & 4C+1S & 13.8 & 786.7641 & 1000 & 6220 & 4.242 & 804.2469 \\ 
8738735 & 693(new) & 2C+1S & 13.9 & 697.8591 & 1000 & 6332 & 4.472 & 893.1428 \\
6191521 & 847\tablenotemark{e} & 1C+2S & 15.2 & 382.9428 & 5000 & 5665 & 4.563 & \nodata \\
6191521 & 847\tablenotemark{e} & \nodata & \nodata & 1489.1858 & 5000 & 5665 & 4.563 & \nodata \\
2162635 & 1032.01 & 1CS+2S & 13.9 & 176.0986 & 4129.9 & 5009 & 3.755 & 1414.9127 \\
2162635 & 1032(new) & \nodata & \nodata & 992.3180 & 1500 & 5009 & 3.755 & 860.8054 \\
2162635 & 1032(new) & \nodata & \nodata & 1351.3427 & 1750 & 5009 & 3.755 & 1219.8301 \\
3230491 & 1096.01 & 1CS &14.7 & 315.33083 & 9592.0 & 5606 & 4.597 & 1275.6733 \\
3218908 & 1108(new) & 3C+1S & 14.6 & 766.6855 & 5000 & 5513 & 4.599 & 824.3164 \\  
10460629 & 1168\tablenotemark{e} &1C+2S & 14.0 & 608.26209 & 22000 & 6449 & 4.232 & \nodata \\
10460629 & 1168\tablenotemark{e} & \nodata & \nodata & 1133.28363 & 22000 & 6449 & 4.232 & \nodata\\
10287723 & 1174.01 & 1CS & 13.5 & 393.5944 & 1474.9 & 4500 & 4.572 & 1197.4066 \\
3962440 & 1208.01 & 1CS & 13.6 & 249.4412 & 3245.9 & 6487 & 4.397 & 1341.5601\\
11342550 & 1421.01 & 1CS & 15.3 & 524.2844 & 9515.3 & 5923 & 4.445 & 1066.7171 \\
10187159 & 1870(new) &  1C+1S & 14.4 & 604.1071 & 6000 & 5185 & 4.440 & 986.8949 \\
5942949 & 2525(new) & 1C+1S & 15.7 & 1326.1614 & 20000 & 4806 & 4.564 & 1179.4970 \\
3241604 & 2824(new) & 1F+1S & 15.3 & 1263.4172 & 5000 & 5881 & 4.516 & 911.0409 \\
1717722 & 3145\tablenotemark{e} & 2C+1S & 15.7 & 1439.1972 & 20000 & 4812 & 4.607 & 1269.6764\\
10284575 & 3210(new) & 1F+1S & 11.9 & 740.72319 & 6000 & 7296 & 4.103 & 850.27786 \\
8636333 & 3349\tablenotemark{e} &  1C+1S & 15.3 & 271.8903 & 1500 & 6247 & 4.489 & 1319.1091 \\
6145201 & 3475(new) & 1F+1S & 13.0 & 789.1 & 1000 & 6517 & 4.382 & \nodata \\
3558849 & 4307\tablenotemark{e} & 1C+1S & 14.2 & 279.9881 & 5000 & 6175 & 4.440 & 1311.0131 \\
4042088 & 6378(new) & 1F+2S &13.4 & 617.65 & 17500 & 6475 & 4.234 & 973.34652 \\
4042088 & 6378(new) & \nodata & \nodata & 661.74 & 4300 & 6475 & 4.234 & 929.2648 \\
9581498 & 7194(new) & 1F+1S & 14.2 & 685.43 & 1000 & 5795 & 4.435 & 905.5850 \\
\hline
\end{tabular}
%} % scalebox ends here
\tablenotetext{1}{Letters in the ``architecture" column have the following meanings: C=planet candidates listed in the KOI catalog as ``CANDIDATE"; F=planet candidates listed in the KOI catalog as ``FALSE POSITIVE"; S=STE we identified; CS=both C and S (i.e. STEs that are already listed in the KOI catalog as planet candidates).}
\tablenotetext{2}{``(new)" after the KOI name indicates the transit-like events that were not reported in the KOI catalog.}
\tablenotetext{3}{Approximate depths from the visual inspection for new candidates.
For the known candidates, catalog values are listed instead.}
\tablenotetext{4}{The $P_{\rm Kep, min}$ is the minimum possible orbital period of the STE candidates determined from the absence of other transits
in the existing {\it Kepler} data. Here we neglect the possibility that other transits fell into the data gaps.}
\tablenotetext{5}{These events are not listed in the KOI catalog,
but have been independently reported in the recent paper by \citet{2015arXiv151202559W}.}
\end{center}
\end{table*}

%%%%%%%%%%%%%%%%%%%%%%%%%%%%%%%%%
%	Table
%	Orbital parameters of STE candidates
%%%%%%%%%%%%%%%%%%%%%%%%%%%%%%%%%
%\include{tableall}
\begin{table*}[!tbh]
%\rotatebox{90}{\begin{minipage}{\textheight}
\begin{center}
\caption{Parameters of the STE Candidates Derived from Our MCMC Analysis\label{tab:list_fit}}
\vspace{0.1cm}
\scalebox{0.77}{
\begin{tabular}{cccccccccccc}
   \hline\hline
KOI & $\rho_\ast$ (g/cc) & $u_1+u_2$ & $u_1-u_2$ & $T_c$ (KBJD) & $P$ (day) & $\cos i$ & $k=R_p/R_\ast$ & $b$\tablenotemark{\dag} & $T$ (hr)\tablenotemark{\dag} & $\tau$ (hr)\tablenotemark{\dag} & $R_p\ (R_\oplus)\tablenotemark{\dag}$\\
\hline
% (1), (2): 1st & 2nd STE
\multicolumn{4}{l}{\it (Uehara Objects of Interest)}\\
847(1)&0.47$\,^{+0.12}_{-0.13}$&0.609$\,^{+0.064}_{-0.063}$&0.166&382.9428$\,^{+0.0049}_{-0.0046}$&930$\,^{+430}_{-380}\tablenotemark{\dag\dag}$&0.00277$\,^{+0.00098}_{-0.00057}$&0.0676$\,^{+0.0021}_{-0.0030}$&0.79$\,^{+0.047}_{-0.12}$&15.69$\,^{+0.34}_{-0.37}$&2.9$\,^{+0.87}_{-1.0}$&5.6$\pm$2.5\\
847(2)&0.48$\,^{+0.12}_{-0.14}$&0.611$\,^{+0.064}_{-0.060}$&0.166&1489.1858$\pm$0.0046&840$\,^{+380}_{-300}\tablenotemark{\dag\dag}$&0.0029$\,^{+0.0011}_{-0.00061}$&0.0680$\,^{+0.0018}_{-0.0022}$&0.776$\,^{+0.048}_{-0.082}$&15.66$\,^{+0.27}_{-0.28}$&2.73$\,^{+0.75}_{-0.72}$&5.7$\pm$2.5\\
% 3145 removed
%3145&1.81$\pm$0.46&0.89$\,^{+0.084}_{-0.33}$&0.465&1439.1972$\,^{+0.0044}_{-0.0040}$&6900$\,^{+2500}_{-1900}$&0.00049$\,^{+0.00017}_{-0.00010}$&0.1500$\,^{+0.0045}_{-0.0031}$&0.802$\,^{+0.048}_{-0.023}$&17.5$\,^{+0.90}_{-3.7}$&8.7$\,^{+3.7}_{-1.7}$&11.7$\pm$1.2\\
1108&0.96$\,^{+0.22}_{-0.29}$&0.68$\pm$0.11&0.106&766.6855$\,^{+0.0031}_{-0.0035}$&1160$\,^{+760}_{-430}$&0.00099$\,^{+0.00039}_{-0.00044}$&0.0665$\,^{+0.0024}_{-0.0019}$&0.46$\,^{+0.18}_{-0.29}$&19.23$\,^{+0.28}_{-0.26}$&1.62$\,^{+0.61}_{-0.33}$&5.5$\pm$1.9\\
671&0.96$\,^{+0.15}_{-0.36}$&0.63$\,^{+0.14}_{-0.13}$&0.0118&786.7641$\,^{+0.0071}_{-0.0074}$&7700$\,^{+2900}_{-2500}$&0.00021$\,^{+0.00012}_{-0.00015}$&0.02675$\,^{+0.00066}_{-0.00062}$&0.29$\,^{+0.24}_{-0.22}$&39.03$\,^{+0.41}_{-0.37}$&1.13$\,^{+0.34}_{-0.096}$&3.9$\pm$2.5\\
693&0.34$\,^{+0.14}_{-0.18}$&0.71$\pm$0.19&0.00304&697.8591$\,^{+0.0078}_{-0.0065}$&980$\,^{+520}_{-470}$&0.0012$\,^{+0.0014}_{-0.00082}$&0.0325$\,^{+0.0012}_{-0.0010}$&0.32$\,^{+0.28}_{-0.24}$&26.58$\,^{+0.62}_{-0.46}$&0.96$\,^{+0.42}_{-0.098}$&3.5$\pm$1.5\\
435&1.10$\,^{+0.094}_{-0.20}$&0.596$\,^{+0.064}_{-0.062}$&0.21&657.2698$\,^{+0.0014}_{-0.0015}$&910$\,^{+210}_{-230}$&0.00152$\,^{+0.00033}_{-0.00012}$&0.0860$\,^{+0.0013}_{-0.0017}$&0.579$\,^{+0.061}_{-0.097}$&15.73$\pm$0.15&2.05$\,^{+0.29}_{-0.30}$&7.8$\pm$3.3\\
2525&3.8$\,^{+1.0}_{-1.1}$&0.64$\,^{+0.28}_{-0.39}$&0.467&1326.1614$\pm$0.0015&1200$\,^{+540}_{-390}$&0.00140$\,^{+0.00052}_{-0.00031}$&0.16$\,^{+0.12}_{-0.0091}$&0.88$\,^{+0.19}_{-0.046}$&5.3$\,^{+1.0}_{-0.17}$&5.12$\,^{+0.19}_{-0.99}$&12.7$\pm$1.3\\
1421$^\ast$&1.40$\pm$0.34&0.61$\,^{+0.12}_{-0.11}$&0.0907&524.2844$\pm$0.0025&2230$\,^{+960}_{-740}$&0.00068$\,^{+0.00025}_{-0.00019}$&0.0913$\,^{+0.0018}_{-0.0031}$&0.54$\,^{+0.092}_{-0.22}$&20.14$\,^{+0.32}_{-0.28}$&2.63$\,^{+0.54}_{-0.62}$&10.2$\pm$4.5\\
 \multicolumn{4}{l}{\it (``Misfit" Singles)}\\
4307&0.90$\,^{+0.73}_{-0.58}$&0.62$\,^{+0.11}_{-0.10}$&0.0329&279.9881$\,^{+0.0023}_{-0.0024}$&610$\,^{+460}_{-400}$&0.0009$\,^{+0.0017}_{-0.00061}$&0.0641$\,^{+0.0013}_{-0.0010}$&0.23$\,^{+0.22}_{-0.17}$&16.85$\pm$0.18&1.14$\,^{+0.23}_{-0.060}$&6.8$\pm$2.6\\
3349&0.73$\,^{+0.26}_{-0.43}$&0.66$\,^{+0.22}_{-0.28}$&0.00906&271.8903$\,^{+0.0086}_{-0.0088}$&510$\,^{+480}_{-290}$&0.0015$\,^{+0.0019}_{-0.00096}$&0.0363$\,^{+0.0020}_{-0.0017}$&0.37$\,^{+0.32}_{-0.28}$&16.48$\,^{+0.43}_{-0.54}$&0.69$\,^{+0.50}_{-0.099}$&3.8$\pm$1.4\\
1870&1.0$\,^{+1.4}_{-0.42}$&0.78$\,^{+0.11}_{-0.10}$&0.329&604.1071$\pm$-0.0015&190$\,^{+270}_{-85}$&0.0014$\,^{+0.0023}_{-0.0011}$&0.0722$\,^{+0.0019}_{-0.0013}$&0.20$\,^{+0.21}_{-0.16}$&11.05$\,^{+0.16}_{-0.14}$&0.83 $\,^{+0.16}_{-0.036}$&7.2$\pm$6.3\\
1208$^\ast$&0.65$\,^{+0.31}_{-0.30}$&0.52$\pm$0.16&$-0.0243$&249.4412$\,^{+0.0015}_{-0.0016}$&65$\,^{+73}_{-36}$&0.0096$\,^{+0.0078}_{-0.0040}$&0.0570$\,^{+0.0016}_{-0.0025}$&0.66$\,^{+0.12}_{-0.40}$&7.29$\,^{+0.14}_{-0.12}$&0.75$\,^{+0.38}_{-0.32}$&7.0$\pm$3.8\\
1174$^\ast$&2.63$\,^{+0.43}_{-0.42}$&0.79$\,^{+0.15}_{-0.21}$&0.564&393.5944$\,^{+0.0038}_{-0.0032}$&310$\,^{+370}_{-88}$&0.00152$\,^{+0.00053}_{-0.00098}$&0.0343$\,^{+0.0029}_{-0.0016}$&0.42$\,^{+0.30}_{-0.31}$&8.98$\,^{+0.23}_{-0.28}$&0.37$\,^{+0.33}_{-0.069}$&2.72$\pm$0.15\\
1096$^\ast$ & 2.68$\,^{+0.36}_{-0.35}$&0.43$\,^{+0.40}_{-0.33}$&0.183&315.3283$\,^{+0.0016}_{-0.0017}$&700$\,^{+190}_{-160}$&0.00268$\,^{+0.00065}_{-0.00041}$&0.20$\,^{+0.26}_{-0.083}$&1.06$\,^{+0.29}_{-0.12}$&3.66$\,^{+0.14}_{-0.10}$&3.66$\,^{+0.13}_{-0.10}$&16.6$\pm$5.9\\
 \multicolumn{4}{l}{\it (EB-like)}\\
1168(1) & 0.558$\,^{+0.036}_{-0.040}$&0.975$\,^{+0.020}_{-0.059}$&0.008801&608.26209$\,^{+0.00042}_{-0.00043}$&130$\,^{+10}_{-9.4}$&0.0155$\,^{+0.0022}_{-0.0017}$&0.42$\,^{+0.12}_{-0.082}$&1.20$\,^{+0.13}_{-0.094}$&4.718$\,^{+0.029}_{-0.046}$&4.718$\,^{+0.029}_{-0.046}$&60$\pm$30\\
1168(2) & 0.548$\,^{+0.040}_{-0.053}$&0.982$\,^{+0.014}_{-0.046}$&0.008801&1133.28363$\pm$0.00040&117$\,^{+11}_{-12}$&0.0168$\,^{+0.0029}_{-0.0023}$&0.43$\,^{+0.14}_{-0.090}$&1.21$\,^{+0.16}_{-0.10}$&4.700$\,^{+0.028}_{-0.038}$&4.700$\,^{+0.028}_{-0.038}$&61$\pm$31\\
% P fixed at the interval of the two (525 days)
%&1168(1)&2.31$\,^{+0.19}_{-0.08}$&0.983$\,^{+0.013}_{-0.044}$&0.008801&608.26209$\,^{+0.00045}_{-0.00044}$&525.02155&0.00378$\,^{+0.00038}_{-0.00031}$&0.44$\,^{+0.15}_{-0.10}$&1.22$\,^{+0.16}_{-0.11}$&4.720$\,^{+0.030}_{-0.037}$&4.720$\,^{+0.030}_{-0.037}$&63$\pm$31\\
%&1168(2)&2.40$\,^{+0.17}_{-0.08}$&0.987$\,^{+0.010}_{-0.032}$&0.008801&1133.28362$\,^{+0.00044}_{-0.00041}$&525.02155&0.00374$\,^{+0.00034}_{-0.00029}$&0.44$\,^{+0.13}_{-0.10}$&1.22$\,^{+0.14}_{-0.11}$&4.702$\,^{+0.028}_{-0.031}$&4.702$\,^{+0.028}_{-0.031}$&63$\pm$31\\
\hline
\end{tabular}
}
\tablenotetext{0}{$^\dag$ Posteriors of these parameters
were derived from those of the fitted parameters.
The error in $R_p$ is based on the posterior of $k$ and
the error in $R_\star$ reported on the CFOP website.}
\tablenotetext{0}{$^{\dag\dag}$ Assuming that the two STEs are 
due to the same object, we obtain $P=1106.243\pm0.007\,\mathrm{days}$
from their interval.}
\tablenotetext{0}{$^\ast$ Prior on the mean stellar density was 
adopted for these systems since they do not host transiting planets
other than the STE candidate.}
\end{center}
%\end{minipage}}
\end{table*}

%%%%%%%%%%%%%%%%%%%%%%%%%%%%%%%%%%%%%
%	Table
%	List of misfit singles
%%%%%%%%%%%%%%%%%%%%%%%%%%%%%%%%%%%%%
\begin{table}[!tbh]
\begin{center}
\caption{List of the ``Misfit" Singles and the Minimum Values of Eccentricity
Required \label{tab:list_misfit}}
\begin{tabular}{cc@{\hspace{0.5cm}}cc@{\hspace{1cm}}cc}
\hline\hline
& & \multicolumn{2}{l}{fiducial values} & \multicolumn{2}{l}{most conservative values}\\
KOI & $P_s$ (days) & $P_{\rm Kep, min}$ (days) & $e_{\rm min}$ &
 $P_{\rm Kep, min}$ (days) & $e_{\rm min}$ \\
\hline
4307 & $610^{+460}_{-400}$			& $1311$ & $0.25$ & $335$ & \nodata \\
3349 & $510^{+480}_{-290}$	& $1319$ & $0.31$ & $532$ & \nodata\\
1870 & $190^{+270}_{-85}$		& $987$	  & $0.50$ & $494$ & $0.31$\\
1208$^\ast$ & $65^{+73}_{-36}$		& $1342$ & $0.77$ & $340$ & $0.50$\\
1174$^\ast$ & $310^{+370}_{-88}$	& $1197$ & $0.42$ & $411$ & \nodata \\
1096$^\ast$ & $700^{+190}_{-160}$	& $1276$ & $0.20$ & $487$ & \nodata\\
%KOI-3210 & \nodata & \nodata & 0.20 & 0.623 & 0.3578 \\
\hline
\end{tabular}
\tablenotetext{0}{Note --- In the ``most conservative values," 
$P_{\rm Kep, min}$ is computed by considering the possibility
that all other transits of the STE candidate are hidden in the data gaps,
although such a possibility is quite low in some cases,
depending on the candidate.}
\tablenotetext{0}{$^\ast$No transiting planet candidates other than
the STE one are known for these systems.}
\end{center}
\end{table}

%%%%%%%%%%%%%%%%%%%%%%%%%%%%%%%%%%%%%
%	Table
%	Occurrence Rate
%%%%%%%%%%%%%%%%%%%%%%%%%%%%%%%%%%%%%
\begin{table}[!tbh]
\begin{center}
\caption{Occurrences of Transiting LPGs \label{tab:occurrence_lpgs}}
\begin{tabular}{cccc}
\hline\hline
 & \# of transiting LPG candidates & total \# of KOIs\tablenotemark{\dag} & fraction of transiting LPG candidates\\
\hline
systems with multiple inner candidates & $4$	& $695$	& $6\times10^{-3}$\\
systems with only one inner candidate	& $2$	& $2963$ &$7\times10^{-4}$\\
\hline
\end{tabular}
\tablenotetext{0}{$^\dag$Objects dispositioned as false positives are not counted as planet candidates.}
\end{center}
\end{table}

\begin{figure*}[htbp]
\begin{center}
\includegraphics[width=\linewidth]{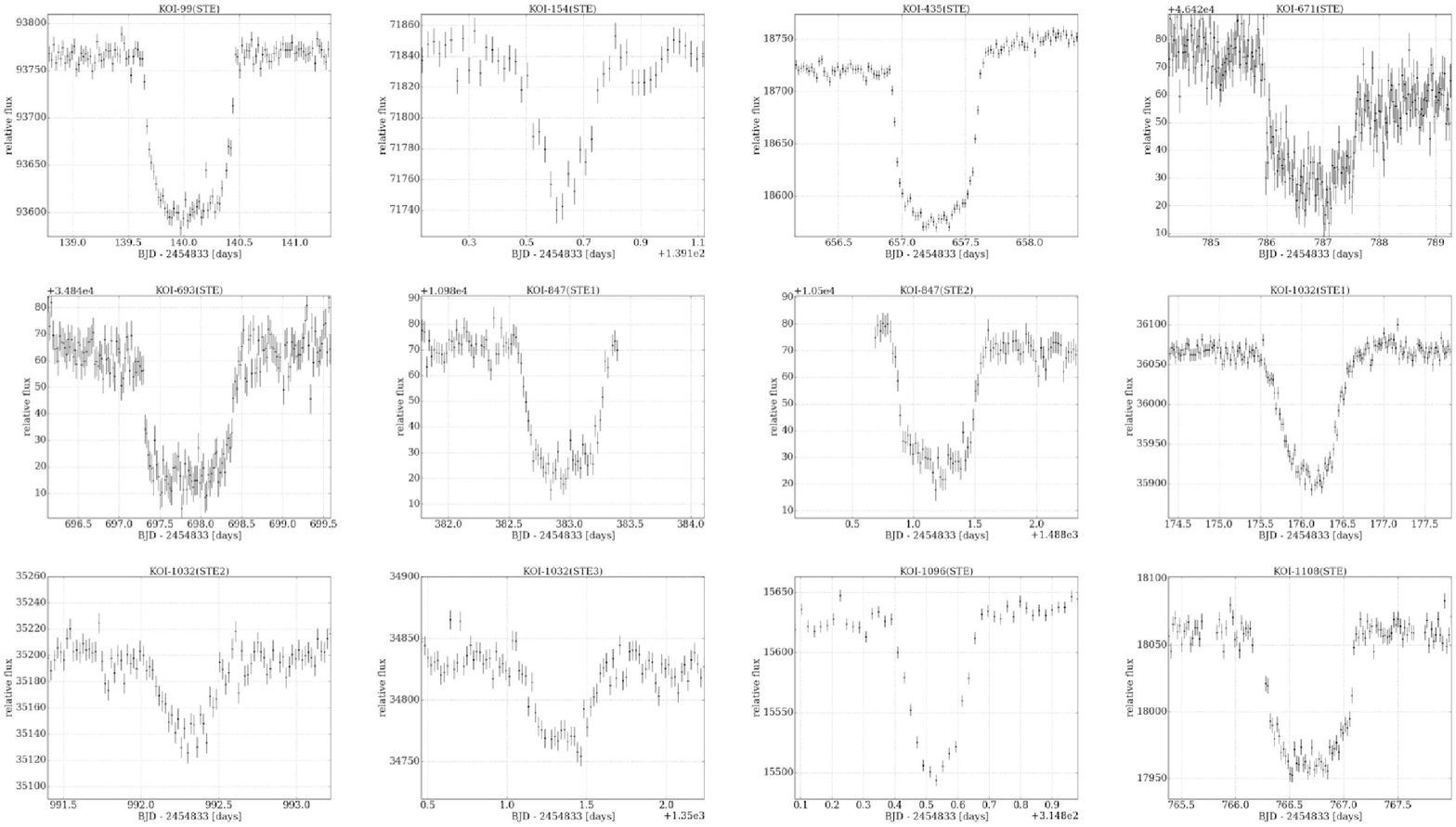}
\includegraphics[width=\linewidth]{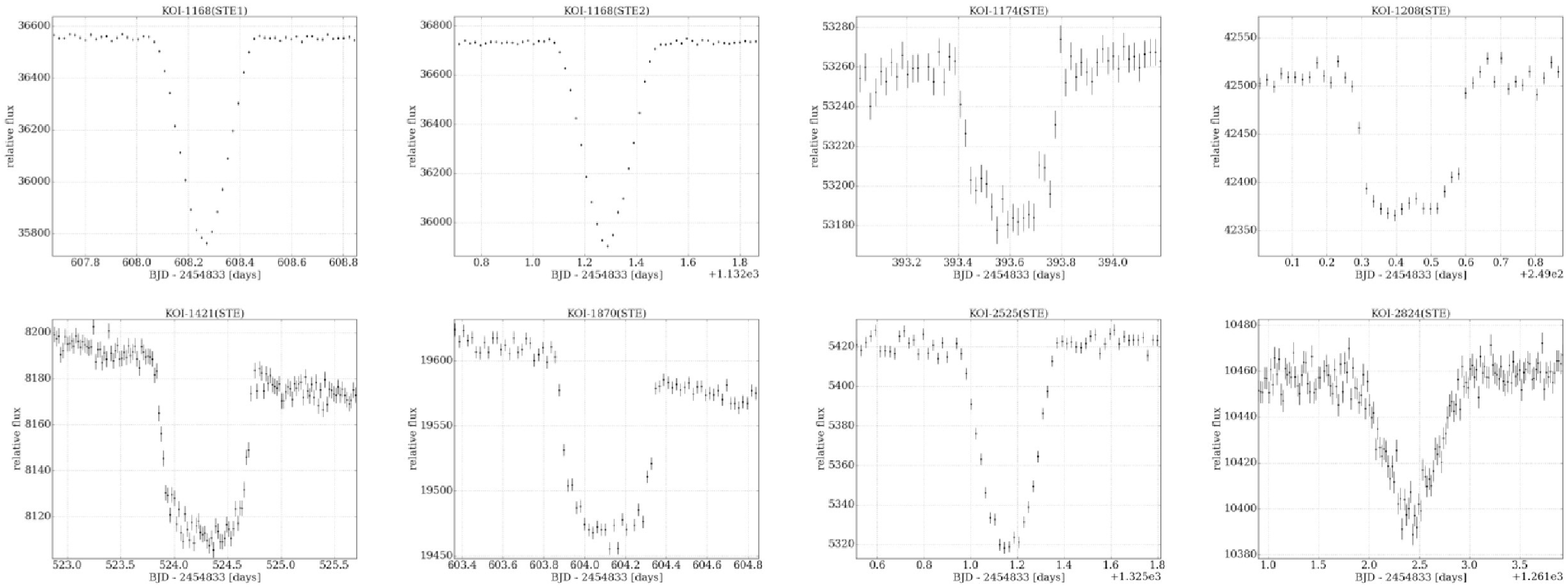}
\includegraphics[width=\linewidth]{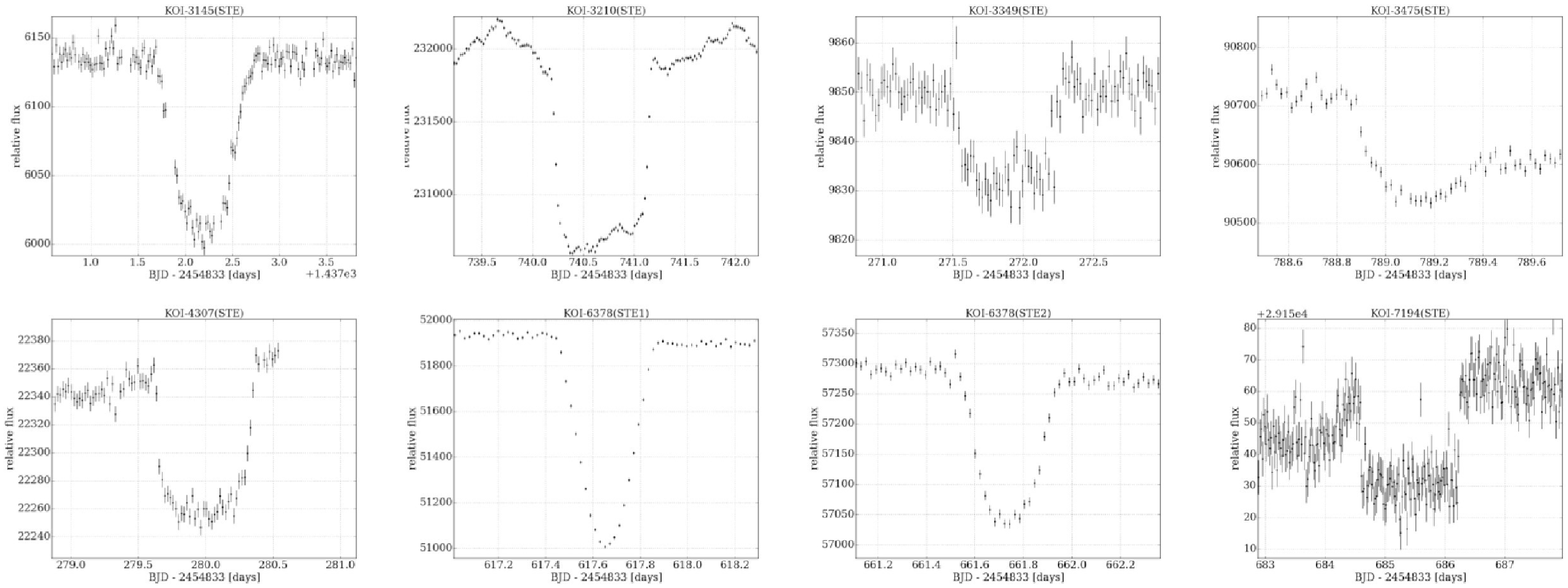}
\caption{PDCSAP light curves of all the 28 STEs in Table \ref{tab:list}
we identified. \label{fig:stes_all}}
\end{center}
\end{figure*}

\begin{figure*}[htbp]
\begin{center}
\includegraphics[width=\linewidth]{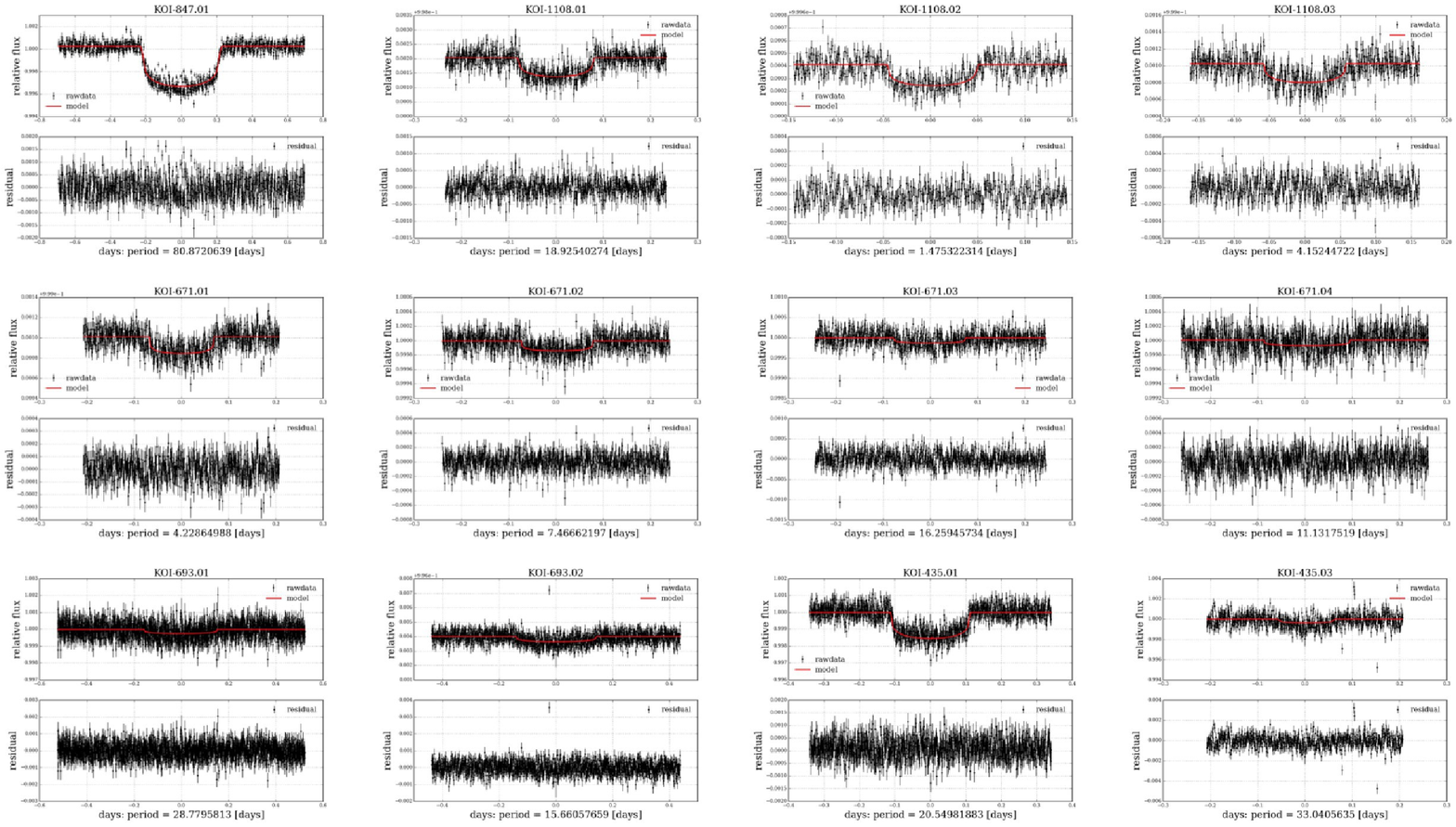}
\includegraphics[width=\linewidth]{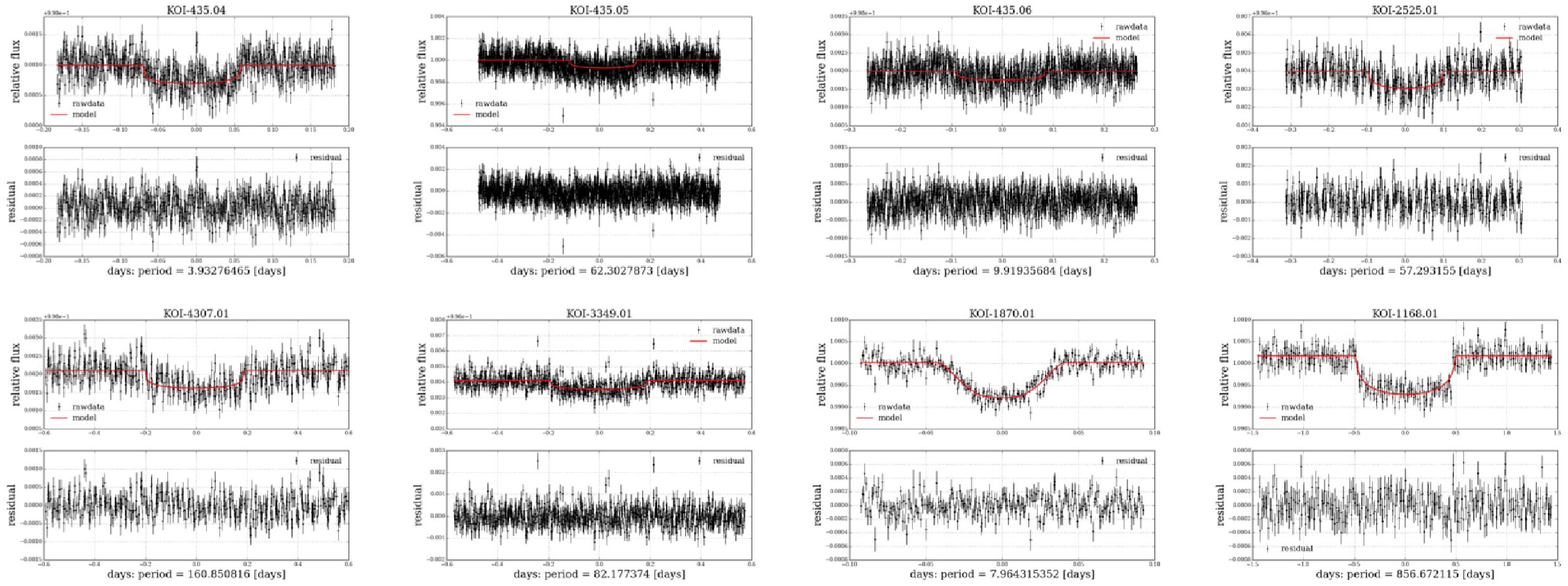}
\caption{Fit to the phase-folded transits of the inner candidates
for 10 of the systems without asterisks in Table \ref{tab:list_fit}.
The black dots with error bars show the binned fluxes 
and the red solid line is the best-fit model obtained from the 
joint fit with the STE candidates. \label{fig:fit_all_inner}}
\end{center}
\end{figure*}

%%%%%%%%%%%%%%%%%%%%%%%%%%%%%%%%%
%	Appendix
%	Light curves of the inner candidates
%%%%%%%%%%%%%%%%%%%%%%%%%%%%%%%%%
\newpage
\appendix
Here we show the PDCSAP light curves of all the 28 STEs we found
(Figure \ref{fig:stes_all})
and the phase-folded transit light curves of the 
inner candidates simultaneously fitted with the 16 STE light curves
in Figure \ref{fig:fit_all} (Figure \ref{fig:fit_all_inner}).
%%%%%%%%%%%%%%%%%%%%%%%%%%%%%%%%%
%	Figure
%	All the STEs
%%%%%%%%%%%%%%%%%%%%%%%%%%%%%%%%%
%%%%%%%%%%%%%%%%%%%%%%%%%%%%%%%%%
%	Figure
%	Inner transits
%%%%%%%%%%%%%%%%%%%%%%%%%%%%%%%%%

\end{document}